\documentclass[aps, pra,10pt,onecolumn,showpacs,citeautoscript,amsmath,amssymb,floatfix,nofootinbib,superscriptaddress,longbibliography]{revtex4-2}

\usepackage{amsfonts}
\usepackage{amssymb}
\usepackage{braket}
\usepackage[utf8]{inputenc}
\usepackage[T1]{fontenc}
\usepackage[colorlinks=true]{hyperref}
\hypersetup{%
  colorlinks = true,
  linkcolor  = black
}
\usepackage{graphicx}
\usepackage[dvipsnames]{xcolor}
\usepackage{soul}

\usepackage{tikz}
\usetikzlibrary{backgrounds}

\usepackage{amsthm}
\usepackage{comment}
\usepackage{cleveref}

\usepackage{enumitem}

\def\be{\begin{equation}}
\def\ee{\end{equation}}
\def\ba{\begin{eqnarray}}
\def\ea{\end{eqnarray}}

\newcommand{\ilcomment}[1]{}
\def\Cl{\mathbb{C}}
\def\Zl{\mathbb{Z}}
\newcommand{\cb}{\mathcal B}

\newcommand{\cd}{\mathcal D}

\newcommand{\cf}{\mathcal F}
\newcommand{\cg}{\mathcal G}
\newcommand{\ch}{\mathcal H}

\newcommand{\calr}{\mathcal R}

\newcommand{\eqa}{\begin{eqnarray}}
\newcommand{\neqa}{\end{eqnarray}}

\usepackage{bm}
\usepackage{bbm}
\def\C{{\mathbbm C}}

\newcommand{\phys}{\mathrm{phys}}

\newcommand{\ketbra}[2] {
	| #1 \rangle \! \langle #2 |}

\def\I{{\mathbb I}}
\def\A{{\sf A}}
\def\B{{\sf B}}
\def\C{{\sf C}}

\def\inv{{\rm inv}}

\DeclareMathOperator{\Tr}{Tr}

\usepackage{geometry}
\begin{document}

\title{Interpreting quantum reference frame transformations through a simple example}

\author{Esteban Castro-Ruiz}
\affiliation{Institute for Quantum Optics and Quantum Information,
Austrian Academy of Sciences, Boltzmanngasse 3, A-1090 Vienna, Austria}
\author{Thomas D. Galley}
\email{thomas.galley@universite-paris-saclay.fr}
\affiliation{Institute for Quantum Optics and Quantum Information,
Austrian Academy of Sciences, Boltzmanngasse 3, A-1090 Vienna, Austria}
\affiliation{Vienna Center for Quantum Science and Technology (VCQ),
Faculty of Physics, University of Vienna, Vienna, Austria}
\affiliation{Université Paris-Saclay, Laboratoire des M{\'e}thodes Formelles, 91190 Gif-sur-Yvette, France}

\author{Leon Loveridge}
\affiliation{Department of Science and Industry Systems, University of South-Eastern
Norway, Kongsberg, 3616, Norway}

\date{September 26, 2025}

\begin{abstract}
Quantum reference frame transformations have been proposed to provide a means by which to translate descriptions of quantum systems relative to each other. At present, there are several differing frameworks for describing quantum reference frames, with concomitantly different transformation rules. Here, we investigate a simple example of three qubit systems with $\mathbb{Z}_2$ symmetry in order to analyze physical and conceptual distinctions between three contemporary approaches to quantum reference frames --- dubbed perspective-neutral, extra particle and operational. By constructing two three-qubit states that are indistinguishable by observables relative to one frame but are distinguishable by  observables relative to another, we show that each of the three approaches provides a distinct transformation rule, which may be understood to reflect differing attitudes towards the global state and the information that can be accessed by each frame. This helps us to shed light on the physical meaning of each approach and the contexts in which they may be most naturally applied.
\end{abstract}

\maketitle

\section{Introduction}
 
\noindent Every description of an experiment typically uses, implicitly or explicitly, a reference frame. In practice, reference frames are concrete, physical objects, subject to the laws of physics. Assuming that all physical systems are ultimately quantum leads to the concept of quantum reference frames (QRFs).
Beginning in earnest with the study of superselection rules
~\cite{PhysRev.88.101,PhysRev.155.1428,wick1970superselection}, quantum reference frames are a rich area of study, with applications to quantum information processing in the absence of shared classical reference frames~\cite{bartlett2007reference}, background independent formulations of quantum theory~\cite{poulinToyModelRelational2006}, the construction of gauge invariant observables in quantum gravity~\cite{Rovelli:1990ph,Rovelli:1990pi}, the paradox of the third particle~\cite{Angelo_2011} and quantum error correction~\cite{PRXQuantum.2.010326,Woods_2020,carrozza2024correspondence}, amongst other topics.

It is natural to ask whether there are quantum generalizations of the well established classical reference frame transformation rules~\cite{arnol2013mathematical,rindler2012essential,aharonov_quantum_1984,Toller1996,Mazzucchi_2001}. The last few years have seen significant efforts in this direction \cite{merriam2005quantum, Angelo_2012,Pereira_2015, Vanrietvelde2020changeof, delaHamette2020quantumreference, Krumm2021quantumreference,H_hn_2021, Hoehn_2022,TemporalLocCastro_Ruiz_2020,glowacki2023operational,castroruiz2023relative, carette2025operational,hohn2019switching, vanrietvelde2023switching, hohn2020switch, hohn2021equivalence, giacomini2019relativistic,fewster2025quantum, streiter2021relativistic, ballesteros2021group, giacomini2021spacetime, mikusch2021transformation, apadula2024quantum, cepollaro2025sum}. Formulations of quantum reference frame transformations have led to insights on the quantum covariance of physical laws \cite{merriam2005quantum, Giacomini_2019, Vanrietvelde2020changeof, hardy2020implementation, giacomini2023einsteinsequivalenceprinciplesuperpositions, de_la_Hamette_2023}, the localizability of events and the connection between QRFs and quantum causal structure \cite{TemporalLocCastro_Ruiz_2020, kabel2024identification, de2024event}, the quantum relativity of subsystems~\cite{Ali_Ahmad_2022, delahamette2021perspectiveneutralapproachquantumframe, castroruiz2023relative, hoehn2023quantumframerelativitysubsystems}, and the relationship between QRFs and edge modes in gauge theory and gravity \cite{carrozza2024edge, kabel2023quantum, araujo2025relational,fewster2025semi}.

Yet, there are alternative formulations of QRF transformations. Are these formulations equivalent? If not, how do they differ, and in which situations should we use each of them? The aim of this work is to shed light on these questions. We compare three different approaches to QRF transformations: the perspective-neutral approach~\cite{Hoehn:2019owq,Vanrietvelde2020changeof,H_hn_2021,delahamette2021perspectiveneutralapproachquantumframe,Hoehn_2022}, the extra particle approach~\cite{castroruiz2023relative} and the operational  approach (e.g. \cite{miyadera2016approximating,loveridge2017relativity,loveridge2018symmetry,glowacki2024quantum,glowacki2024quantum1,jorquera2025uncertainty} for the general framework, \cite{glowacki2023operational,carette2025operational} for the transformation rule). In these three approaches, there is a global state from which different `QRF viewpoints' can be derived. 

We study a simple example in which two different global states are compatible with the same state relative to one reference frame, but yield different states relative to another. This raises the problem of finding whether it is possible to deduce the state relative to the second QRF, given the information available to the first QRF. We show that the different QRF transformation approaches give different answers to this problem, and so they may be regarded as physically inequivalent. Furthermore, we argue that the approaches differ on what physical quantities are assumed to be accessible, either in principle or in practice, and discuss the physical situations to which each approach may be most naturally applied.

To keep technicalities at a minimum, we consider a 3-partite system with $\mathbb{Z}_2$ as a symmetry group. This allows us to focus on the conceptual questions regarding the physical interpretation of QRF transformations, clarifying their operational meaning. We will see that the three frameworks have different conceptual underpinnings and domains of application, and as such the example we introduce might not be an equally natural physical scenario for the three frameworks to be applied to. Nevertheless, we believe that the benefit of applying all three frameworks to a single simple physical situation provides concrete insights into some of the important practical differences between them. In Section \ref{sec:possible_resolutions} we provide outlines of the three different frameworks which include more general considerations, including motivations which are native to them and are not directly related to the scenario we introduce.  Moreover, we hope that the concrete situation we analyze makes our paper accessible and interesting to both newcomers and experts in this field.\footnote{The example we discuss bears similarities to the argument presented in ~\cite[II, B,3]{hoehn2023quantumframerelativitysubsystems} which contrasts approaches to QRFs based on the incoherent twirl to the perspective neutral which is based on the coherent twirl. Moreover, as in our case, the invertibility (or not) of the `reduction map' (see~\Cref{{mainargument}} of the present paper) plays an important role in differentiating the different approaches. }

\section{The problem}

One of the main aims of the subject of QRF transformations is to address the following question:

\medskip
\medskip

\tikzstyle{background rectangle}=[thin,draw=black]
\begin{tikzpicture}[show background rectangle]

\node[align=justify, text width=0.85\textwidth, inner sep=1em]{
In the absence of an external reference frame, how are states relative to internal quantum reference frames described and how are these relative states related to one another?
};

\node[xshift=3ex, yshift=-0.7ex, overlay, fill=white, draw=white, above 
right] at (current bounding box.north west) {
\textit{Question }
};

\end{tikzpicture}

We will address this question by considering quantum reference frames to be quantum systems (with certain properties) which single out a subset of observables. For example, in the case of translation symmetry, a meaningful observable would be relative position with respect to one of the systems. States relative to a quantum reference frame are the states assigned by an external observer with access to the observables singled out by the corresponding frame.

Let us now consider a specific example in which we may address  the above boxed question. Suppose that two observers, Alice and Bob, describe a quantum system $\sf C$ in the absence of an external reference frame for the group $\Zl_2$. We assume that Alice and Bob have access to an internal QRF (i.e., may measure observables relative to their QRF), which we call $\sf A$ and $\sf B$, respectively. We moreover assume there is a global state describing the total system, which is independent of the external frame. 

Given the state of $\sf{BC}$ relative to Alice's reference frame, $\sf{A}$, what can she say about the state of $\sf{AC}$ relative to Bob's reference frame, $\sf B$? We will see that the answer to 
this question is not unique and will depend on what information the observers have. Moreover, we will see that changes of QRF proposed in the literature are not the same. 
The reason is twofold i) different approaches do not agree on which observables are measurable (either in principle or in practice),  ii) changes of QRFs play different roles -- they relate the descriptions relative to different QRFs given a constraint on the physical variables but they are also used to describe what one observer can infer about another observer's description. 
\begin{figure}
    \centering
    \includegraphics[width=
    \linewidth]{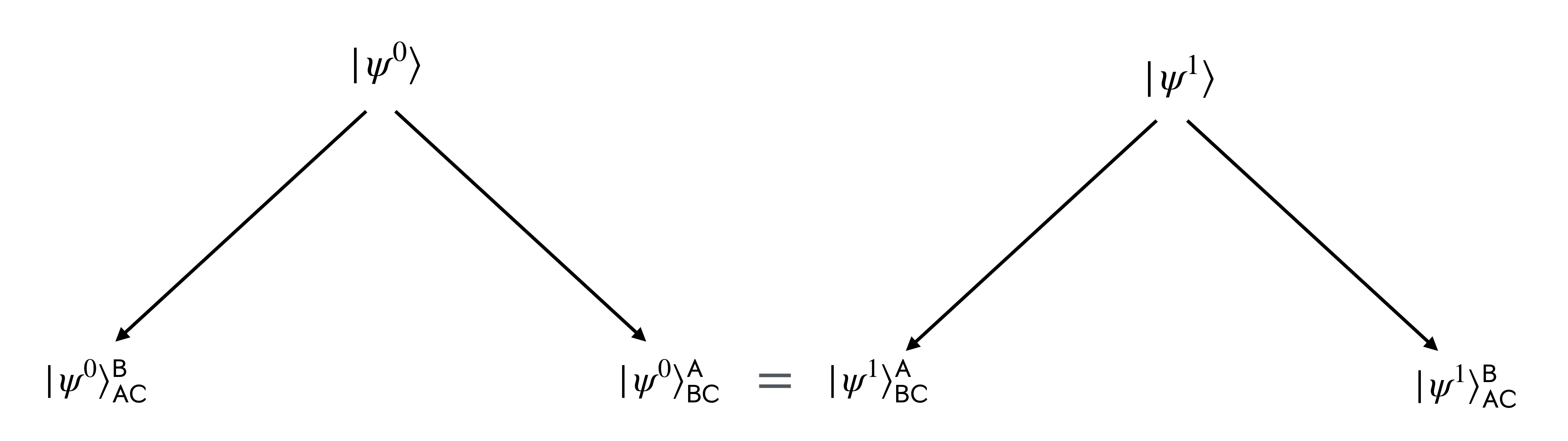}
    \caption{This diagram illustrates the main argument. Two global invariant states $\ket{\psi^0}$ and $\ket{\psi^1}$ lead to the same state $\ket{\psi^0}_{\B\C}^\A = \ket{\psi^1}_{\B\C}^\A$ relative to Alice but different states $\ket{\psi^0}_{\A\C}^\B$ and $\ket{\psi^1}_{\A\C}^\B$ relative to Bob. This entails that there is no unique prescription for mapping from the state relative to Alice to the state relative to Bob.}
\label{fig:outline_of_main_arg}
\end{figure}

\subsection{Setup of the scenario}

Let us consider a classical configuration space $M = \{0,1\}$ corresponding to two positions that a classical system $\sf C$ can take; these are the two pure states of $\sf C$.  The group $\Zl_2$, which consists of the identity e and the finite  translation f, acts as on $M$ as:
\begin{align}
    e &: i \mapsto i , \\
    f &: i \mapsto i \oplus 1, 
\end{align}
where $\oplus$ denotes addition modulo 2.

An external reference frame in particular fixes the origin of the configuration space to be $0$ or $1$. In the absence of an external reference frame all states related by a translation are equivalent for Alice and Bob: the total system is described as translationally invariant. 

The configuration space $M$ is that of a classical bit, and its quantum analogue is a qubit $\Cl^2 \simeq L^2(M)$. The states $\ket 0$ and $\ket 1$ of the qubit correspond to the qubit being in a `classical' state -- a state with a definite position, 0 or 1. The action of the translation group $\Zl_2$ is given by a representation in $\mathbb{C}^2$, chosen to be:
\begin{align}
    e &\mapsto \I, \\
    f &\mapsto X,
\end{align}
where $\I$ is the identity and $X$ acts as:
\begin{align}
    X \ket i = \ket{i \oplus 1} .
\end{align}
This unitary representation is reducible and decomposes into a direct sum of a trivial and a sign representation (which are both necessarily one dimensional). A basis for the trivial representation is $\ket + = \frac{\ket 0 + \ket 1}{\sqrt 2}$  and a basis for the sign representation is $\ket - = \frac{\ket 0 - \ket 1}{\sqrt 2}$. We have:
\begin{align}
    X^i \ket + &= \ket +, \\
    X^i \ket - &= (-1)^i \ket -,
\end{align}
where $X^0 = \I$.

Although $\ket -$ is not invariant under the representation of $\Zl_2$ as a vector,  it is however invariant as a quantum state (in the sense of a ray in Hilbert space or a density operator). 
More generally, any density operator $\rho$ which is diagonal in the $\{\ket{+},\ket{-}\}$ basis is invariant: $\rho = X^i \rho (X^i)^\dagger$, for $\rho = p \ketbra{+}{+} + (1-p) \ketbra{-}{-}$.

We now introduce quantum reference frames; these are quantum systems relative to which the state of the systems of interest can be expressed. The procedure by which this is done will be made explicit in the following.

Consider two systems $\A$ and $\B$, serving as quantum reference frames, so that the total Hilbert space is $\ch = \ch_\A \otimes \ch_\B \otimes \ch_\C$. Each tensor factor is a qubit, i.e., is isomorphic to $\mathbb{C}^2$. The representation of the translation group $\Zl_2$  is given by the following action on the three qubits:
\begin{align}
    e: \ket \psi_{\A\B\C} &\mapsto \I_\A \otimes \I_\B \otimes \I_\C \ket{\psi}_{\A\B\C} \\ 
    f: \ket \psi_{\A\B\C} &\mapsto X_\A \otimes X_\B \otimes X_\C \ket{\psi}_{\A\B\C}.
\end{align}
We write the unitaries in this representation of $\Zl_2$ as $X^i_\A \otimes X^i_\B \otimes X^i_\C$ for $i = 0,1$. Because the action of $\Zl_2$ on $\ch$ is compatible with the view of a (in fact any) potential external frame, $\ch$ and the $\Zl_2$ action on it are said to be represented in a `global' perspective. 

Under the  given representation of $\Zl_2$, $\ch$ decomposes as $\ch = \ch_0 \oplus  \ch_1$, where $\ch_0 \cong \mathbb{C}^4$ carries the trivial representation and $\ch_1 \cong \mathbb{C}^4$ carries the sign representation. 
$\frac{1}{\sqrt 2}(\ket{000} + \ket{111})$ is an example of a state lying in the trivial representation since $(X_\A \otimes X_\B \otimes X_\C)(\frac{1}{\sqrt 2}(\ket{000} + \ket{111})) = \frac{1}{\sqrt 2}(\ket{000} + \ket{111})$.

A general vector in the trivial subspace $\ch_0$ is of the form
\begin{align}
    \alpha (\ket{000} + \ket{111}) + \beta (\ket{001} + \ket{110}) + \gamma (\ket{010} + \ket{101}) + \delta (\ket{100}+ \ket{011}), \label{eq:triv_sub}
\end{align}
and a vector in the sign subspace $\ch_1$ is of the form:
\begin{align}
    \alpha (\ket{000} - \ket{111}) + \beta( \ket{001} - \ket{110}) + \gamma (\ket{010} - \ket{101}) + \delta (\ket{100}- \ket{011}).\label{eq:sign_sub}
\end{align}

\subsection{Main argument}\label{mainargument}

Consider the two unit vectors:\footnote{In the following we will not always explicitly distinguish between states and (unit) vectors when the context is obvious.}
\begin{align}
    \ket{\psi^0} = \frac{1}{2} \left((\ket{001} + \ket{110}) - (\ket{100} + \ket{011}) \right) \in \ch_0 ,\label{globalzero}\\
    \ket{\psi^1} = \frac{1}{2} \left((\ket{001} - \ket{110}) + (\ket{100} - \ket{011}) \right) \in \ch_1 .\label{globalone}
\end{align}
$\ket{\psi^0}$ and $\ket{\psi^1}$ define invariant global states, the first lying in the trivial and the second in the sign subspace. Invariance of these states implies that they are independent of an external reference frame for discrete translations $\Zl_2$.

We denote by $\calr_{\bra{0}_\A}$ the map $\ch \mapsto \ch_\B \otimes \ch_\C$, $\calr_{\bra{0}_\A}(\ket \psi) = \sqrt{2} \bra 0_A \ket{\psi}$ (and similarly for $\calr_{\bra{0}_\B}$), where the constant $\sqrt{2}$ ensures that the output state is normalised. Then, the state of $\sf{BC}$ relative to $\sf A$ may be defined by $\ket{\psi}_{\sf{BC}}^{\sf A} = \mathcal{R}_{\bra{0}_{\sf A}} \ket{\psi}$, which is interpreted as the state of $\B\C$ when $\A$ is `at the origin' (see \cite{carette2025operational} and Appendix A below for an operational justification of this interpretation), i.e. in the state $\ket 0 _\A$.\footnote{Technically one first defines a set of $\Zl_2$ coherent states for $\A$: \{$\ket i_\A = X^i \ket 0_\A\}_{i = 0,1}$ and then conditions on the seed (or reference) state $\ket 0_\A$.} Physically, the vectors $\ket{ij}^{\sf A}_{\sf{BC}}$ correspond to states where $\sf{B}$ and $\sf{C}$ are in positions $i$ and $j$, respectively, with respect to Alice's QRF, $\sf{A}$.

The states $\ket{\psi^0}_{\B\C}^\A$ and $\ket{\psi^1}_{\B\C}^\A$ relative to Alice are
\begin{align}
    \ket{\psi^0}_{\B\C}^\A = \calr_{\bra{0}_\A}(\ket{\psi^0}) = \frac{1}{\sqrt{2}} (\ket{01} - \ket{11}), \label{zerowrtA} \\
    \ket{\psi^1}_{\B\C}^\A = \calr_{\bra{0}_\A}(\ket{\psi^1}) = \frac{1}{\sqrt{2}} (\ket{01} -\ket{11}). \label{onewrtA}
\end{align}
Thus both $\ket{\psi^0}$ and $\ket{\psi^1}$, in the global perspective, give rise to the same reduced state for $\B\C$ as described by $\A$. What are the reduced states relative to $\B$? These are given by
\begin{align}
    \ket{\psi^0}_{\A\C}^\B = \calr_{\bra{0}_\B}(\ket{\psi^0}) =  \frac{1}{\sqrt{2}} (\ket{01} - \ket{10}), \label{statefrombob0} \\
    \ket{\psi^1}_{\A\C}^\B =  \calr_{\bra{0}_\B}(\ket{\psi^1}) = \frac{1}{\sqrt{2}} (\ket{01} + \ket{10}). \label{statefrombob1}
\end{align}
Thus the two global states give different (indeed orthogonal) states of $\sf{AC}$ as seen from Bob's perspective. 

Hence, in general there can be two different global states $\ket{\psi^0}$ and $\ket{\psi^1}$ which give the same states relative to Alice: $ \ket{\psi^0}_{\B\C}^\A = \ket{\psi^1}_{\B\C}^\A$, but however give different states relative to Bob: $\ket{\psi^0}_{\A\C}^\B  \neq \ket{\psi^1}_{\A\C}^\B$. See~\Cref{fig:outline_of_main_arg} for an illustration of the scenario.

This observation means that the two global states give the same expectation values for all observables relative to $\A$ but different expectation for observables relative to $\B$ (for a definition of relative observables, see~\Cref{sec:observable_based}). More concretely, there exists an observable $T_{\sf{AC}}$ relative to $\sf B$ such that $\bra{\psi^0}_{\A\C}^\B T_{\sf{AC}} \ket{\psi^0}_{\A\C}^\B \neq \bra{\psi^1}_{\A\C}^\B T_{\sf{AC}} \ket{\psi^1}_{\A\C}^\B$. Note that we compare the expectation value of different states with the same observable. This implies a genuine physical difference between the predictions made with $\ket{\psi^0}_{\A\C}^\B$ and those made with $\ket{\psi^1}_{\A\C}^\B$.

\subsection{Paradox}

A key object in recent work on quantum reference frames is the change of quantum reference frame map, which transforms a state relative to one QRF to a state relative to another. The approaches discussed here all begin with a global state from which the relative states are derived, but we note that some approaches  adopt the state relative to one observer as a starting point ( e.g. ~\cite{Giacomini_2019,delaHamette2020quantumreference}).

In the example given above, the state relative to Alice is given by $\ket{\psi}^\A_{\B\C} = \frac{1}{\sqrt{2}}(\ket{01} - \ket{11})$. Given only this state (`Alice's perspective') what can be said, if anything, about the state $\ket{\psi}^\B_{\A\C}$ from Bob's perspective? In the approaches that include the global state, this will depend on the global state $\ket{\psi}_{\A\B\C}$ from which the state relative to Alice is derived. However, as can be seen from the example, there is no unique global state consistent with Alice's relative state $\ket{\psi}^\A_{\B\C}$: both $\ket{\psi_0}$ and $\ket{\psi_1}$ are consistent with Alice's reduced description $\ket{\psi}^\A_{\B\C}$. Moreover, $\ket{\psi_0}$ and $\ket{\psi_1}$ give rise to different states relative to Bob, hence there is no unique state $\ket{\psi}^\B_{\A\C}$ consistent with Alice's relative state $\ket{\psi}^\A_{\B\C}$.

Thus the following problem arises:

\tikzstyle{background rectangle}=[thin,draw=black]
\begin{tikzpicture}[show background rectangle]

\node[align=justify, text width=0.85\textwidth, inner sep=1em]{
Given that a relative description of $\B$ and $\C$ with respect to Alice is consistent with multiple relative descriptions of $\A$ and $\C$ relative to Bob, how, if it is possible, can we define a change of reference frame from Alice to Bob?
};

\node[xshift=3ex, yshift=-0.7ex, overlay, fill=white, draw=white, above 
right] at (current bounding box.north west) {
\textit{Problem }
};

\end{tikzpicture}

Thus, given that Alice's description of $\B\C$ is given by  $\ket{\psi}_{\B\C}^\A = \ket{\psi^0}_{\B\C}^\A = \ket{\psi^1}_{\B\C}^\A$ two possibilities present themselves:

\tikzstyle{background rectangle}=[thin,draw=black]
\begin{tikzpicture}[show background rectangle]

\node[align=justify, text width=0.85\textwidth, inner sep=1em]{
\begin{enumerate}
\vspace*{-\baselineskip}
    \item Alice cannot uniquely determine Bob's state assignment to be either $\ket{\psi^0}_{\A\C}^\B$ or $\ket{\psi^1}_{\A\C}^\B$.
    \begin{enumerate}
        \item There does not exist a change of reference frame map.
        \item Alice can only have partial knowledge of the state relative to Bob; one introduces a change of reference frame operator which is partial: it maps from Alice's description to what she can infer about Bob's description. 
    \end{enumerate}
    \item  Alice can uniquely determine Bob's state assignment to be either $\ket{\psi^0}_{\A\C}^\B$ or $\ket{\psi^1}_{\A\C}^\B$.
    \begin{enumerate}
        \item Restrict the set of global states so that they are in one-to-one correspondence with the states relative to Bob. Alice and Bob are aware of this restriction.
        \item Include the remaining invariant degrees of freedom into the description,  such that the states relative to any QRF are in one-to-one correspondence with the global states.
    \end{enumerate}
\end{enumerate}

};

\node[xshift=3ex, yshift=-0.7ex, overlay, fill=white, draw=white, above 
right] at (current bounding box.north west) {
\textit{Resolutions }
};

\end{tikzpicture}

These different possibilities are implemented by the following frameworks:

\medskip
\medskip

\tikzstyle{background rectangle}=[thin,draw=black]
\begin{tikzpicture}[show background rectangle]

\node[align=justify, text width=0.85\textwidth, inner sep=1em]{

\vspace*{-\baselineskip}
\begin{enumerate}
    \item 
\begin{enumerate}

    \item \textbf{Trivial approach: } No change of reference frame map.
    \item\textbf{Operational approach}~\cite{carette2025operational}: Alice considers both $\ket{\psi^0}_{\A\C}^\B$ and $\ket{\psi^1}_{\A\C}^\B$ as valid states relative to Bob. She can group them into an equivalence class $[\ket{\psi^0}_{\A\C}^\B] = [\ket{\psi^1}_{\A\C}^\B]$ in a natural way based on her choice of frame, and define a change of QRF: $\ket{\psi^0}^{\A}_{\B\C} \mapsto [\ket{\psi^0}_{\A\C}^\B]$.\footnote{The equivalence classes arise through fixing a frame observable; see below and \cite{carette2025operational,glowacki2023operational}. Moreover, an invertible map between the convex sets of the thus defined classes can be obtained by also imposing the  analogous equivalence relation on the left hand side of the frame change.}
    \end{enumerate}
    \item
    \begin{enumerate}
        \item \textbf{Perspective neutral approach}~\cite{delahamette2021perspectiveneutralapproachquantumframe}: the global invariant state is constrained. It lies in $\ch_0$, known as the physical subspace. This ensures that the map $\bra{0}_\A : \ch_0 \to \ch_\B \otimes \ch_\C$ is invertible, so that there exists a unique state relative to Bob. The change of QRF is therefore $\ket{\psi^0}^{\A}_{\B\C} \mapsto \ket{\psi^0}_{\A\C}^\B$.
        \item \textbf{Extra-particle approach}~\cite{castroruiz2023relative}:  Alice can in principle determine which irreducible subspace $\ch_0$ or $\ch_1$ the global invariant state lies in. This information is encoded in an `extra-particle' relative to Alice, $\overline{\sf{BC}}$, which together with her relative state gives her enough information to reconstruct Bob's state assignment, 
        allowing her to discriminate between the two possible global states $\ket{\psi^0}$ and $\ket{\psi^1}$.The change of QRF is thus either  $\ket{0}^{\sf A}_{\overline{\sf{BC}}} \ket{\psi^0}^{\A}_{\B\C} \mapsto \ket{0}_{\overline{\sf{AC}}}^\B \ket{\psi^0}_{\A\C}^\B$ or $\ket{1}_{\overline{\B\C}}^\A \ket{\psi^1}^{\A}_{\B\C} \mapsto \ket{1}_{\overline{\B\C}}^\A \ket{\psi^1}_{\A\C}^\B$ depending on the global state.
    \end{enumerate}
\end{enumerate}

};

\node[xshift=3ex, yshift=-0.7ex, overlay, fill=white, draw=white, above 
right] at (current bounding box.north west) {
\textit{Implementation }
};

\end{tikzpicture}

\medskip

A more detailed account of each resolution, which includes the motivations and conceptual backgrounds of the different frameworks, is given in~\Cref{sec:possible_resolutions}.

\section{Constructing global invariant states and observables}

Before discussing each QRF approach in more detail, it is useful to describe some techniques to construct global invariant states. As noted above, the unit vectors $\ket{\psi^0}$ and $\ket{\psi^1}$  both describe global states (as rank-$1$ projectors), i.e., states describing the three systems of interest, which are invariant under the symmetry group. In contradistinction, $\ket{\psi^1}$ is not invariant as a vector, since $X_\A \otimes X_\B \otimes X_\C  \ket{\psi^1} = - \ket{\psi^1}$.

The general invariant states, i.e., those that satisfy $\rho = (X_\A^i \otimes X_\B^i\otimes X_\C^i) \rho (X_\A^i \otimes X_\B^i\otimes X_\C^i)$ for all $i$, are block diagonal: $\rho \in \cb(\ch_0) \oplus \cb(\ch_1)$. An invariant state $\rho_\inv$ can be constructed from a non-invariant state $\rho$ via a ``$G$-twirl'' (sometimes known as an incoherent or weak twirl):
\begin{align}
    \cg&: \cb(\ch_\A \otimes \ch_\B \otimes \ch_\C) \to \cb(\ch_\A \otimes \ch_\B \otimes \ch_\C)^G \simeq \cb(\ch_0) \oplus \cb(\ch_1) ,\\
    \cg&: \rho \mapsto \frac{1}{2} \sum_{i = 0,1} (X_\A^i \otimes X_\B^i\otimes X_\C^i) \rho (X_\A^i \otimes X_\B^i\otimes X_\C^i) . 
\end{align}
The index $i \in \{0,1\}$ which labels the spaces $\cb(\ch_i)$ is also known as the \emph{charge}, and $\cb(\ch_i)$ the corresponding \emph{charge sector}. We write a superscript `$G$' above to denote the invariants, keeping in mind that $G$ acts through the given unitary representation.

The global invariant states in the extra-particle and the operational approach are of the form $\rho_\inv$ and can be obtained via the weak twirl, which is surjective onto the set of invariant states. By contrast, the global invariant pure states in the perspective neutral approach $\ket{\psi}_\phys$ are constrained to lie fully in the $\ch_0$ subspace, i.e. they are required to be invariant \emph{as vectors}. A perspective neutral state $\ket{\psi}_\phys$ can be constructed from a general pure state $\ket \psi$ via the projection onto the $\ch_0$ subspace:
\begin{align}
    \Pi_0&: \ch_\A \otimes \ch_\B \otimes \ch_\C \to \ch_0, \\
    \Pi_0&: \ket \psi \mapsto \frac{1}{2} \sum_{i = 0,1} (X_\A^i \otimes X_\B^i\otimes X_\C^i) \ket \psi . 
\end{align}
This `coherent group average' $\Pi_0$ is sometimes known as the strong or coherent $G$-twirl.

The incoherent group average $\cg$ is a completely positive trace preserving map, corresponding for example to a non-selective measurement of the total charge or a form of ignorance about which of the two transformations $\{X_\A^i \otimes X_\B^i\otimes X_\C^i\}_{i = 0,1}$ was applied. On the other hand the map $\rho \mapsto \Pi_0 \rho \Pi_0$ is completely positive but not trace-preserving, and could be interpreted as a form of post-selection of a total charge measurement. Alternatively, it can be viewed purely abstractly as a tool for constructing the set of physical states, as will be discussed in~\Cref{sec:PN}.

\section{Possible resolutions}\label{sec:possible_resolutions}

\subsection{Perspective-neutral approach}\label{sec:PN}

\subsubsection{The physical Hilbert space and the reduction map}

One way out of the problem raised is to observe that if we impose that the global pure state of the three particles must lie fully in $\ch_0$, then given Alice's reduced state $\frac{1}{\sqrt{2}}(\ket{01}_{\B\C}^\A-\ket{10}_{\B\C}^\A)$ there is a unique global pure state, here $\ket{\psi^0}$, which is compatible with her description, i.e., whose image under $\calr_{\bra 0_\A}$ is Alice's reduced state written above. The map $\calr_{\bra 0_\A}: \ket{\psi^0} \mapsto \ket{\psi^0}^\A_{\B\C} := \sqrt 2  \bra{0}_{\A} \ket{\psi^0}$ is invertible when restricted to the subspace $\ch_0$, with inverse given by $\ket{\psi^0}^\A_{\B\C} \mapsto  {\sqrt 2} \Pi_0 (\ket 0_\A \ket{\psi^0}^\A_{\B\C}) = \ket{\psi^0}$.
The invertibility of the reduction map $\calr_{\bra 0_\A}$ entails that one can define a unitary change of quantum reference frame $\ket{\psi^0}^\A_{\B\C} \mapsto \ket{\psi^0} \mapsto \ket{\psi^0}^\B_{\A\C}$. 

This resolution is implemented in the perspective neutral framework, which takes inspiration from Dirac quantisation of constrained systems, in which physical (pure) states are given as those unit vectors that lie in the kernel of the quantised constraint operator, or equivalently as the space of vectors invariant under the group generated by the constraint.  The subspace $\ch_0$ is known as the \emph{physical Hilbert space}. Pure states which lie in this subspace are known as \emph{physical states} or \emph{perspective neutral states}. The state $\ket{\psi^1}$ which transforms non-trivially as a vector under $(X_\A \otimes X_\B \otimes X_\C)$ is therefore not a physical state, and is excluded.

The perspective neutral framework provides an interpretation of changes of quantum reference frames as `changes of quantum coordinates', as explored below. For more details of the perspective neutral framework we refer the reader to~\cite{delahamette2021perspectiveneutralapproachquantumframe}  which contains a  general account of the framework, and applies to unimodular Lie groups (including discrete groups) with non-ideal reference frames.

\subsubsection{QRFs as `quantum coordinates' and the physical equivalence of different perspectives}

The different choices of QRF correspond to different choices of `quantum coordinates' being used to describe the same physical state. In our example this means that the states $\ket{\psi^0}^\A_{\B\C}$ and $\ket{\psi^0}^\B_{\A\C}$ are two different ways of representing the same state $\ket{\psi^0}$. This is analogous to choosing a coordinate system placing particle $\A$ at the origin or particle $\B$ at the origin in classical mechanics. The invertibility of the change of quantum reference frame map is thus a crucial feature of this framework, since it is necessary for the interpretation as a change of quantum coordinates.

The physical equivalence of the states $\ket{\psi^0}^\A_{\B\C}$,  $\ket{\psi^0}^\B_{\A\C}$ and $\ket{\psi^0}$ can be seen as follows. Physical observables are self-adjoint operators $\hat{A}_0$ with support on $\ch_0$, i.e. which obey:
\begin{align}
    \hat A_0 \Pi_0 = \Pi_0 \hat A_0 .
\end{align}
From the equality $\ket{\psi^0} = {\sqrt{2}} \Pi_0 (\ket 0_\A \ket{\psi^0}^\A_{\B\C}) = {\sqrt{2}}  \Pi_0 (\ket 0_\B \ket{\psi^0}^\B_{\A\C})$ it follows that:
\begin{align}
    \bra{\psi_0} \hat A_0 \ket{\psi_0} ={ 2} \big(\bra 0_\A \bra{\psi^0}^\A_{\B\C}\big) \hat A_0 \big( \ket 0_\A \ket{\psi^0}^\A_{\B\C}\big) = { 2}\big(\bra 0_\B \bra{\psi^0}^\B_{\A\C}\big) \hat A_0 \big(\ket 0_\B \ket{\psi^0}^\B_{\A\C}\big) . 
\end{align}
The {vectors} $ { \sqrt{2}}\ket 0_\A\ket{\psi^0}^\A_{\B\C}$,  ${ \sqrt{2}} \ket 0_\B\ket{\psi^0}^\B_{\A\C}$ and $\ket{\psi^0}$ give the same expectation values  on all physical observables $\hat A_0 \in \cb(\ch_0)$, and are as such physically equivalent.

We stress that the equality $\big(\bra 0_\A \bra{\psi^0}^\A_{\B\C}\big) \hat A_0 \big( \ket 0_\A \ket{\psi^0}^\A_{\B\C}\big) = \big(\bra 0_\B \bra{\psi^0}^\B_{\A\C}\big) \hat A_0 \big(\ket 0_\B \ket{\psi^0}^\B_{\A\C}\big)$ is not the same as the (trivial) case $\bra \psi \hat A \ket \psi = \bra \psi U^\dagger (U \hat A U^\dagger) U \ket \psi$ corresponding to a general invariance of probabilities under a unitary transformation \emph{of both state and observable} (or equivalently a  freedom in a choice of basis when computing probabilities in quantum theory). Rather it witnesses a form of invariance or relativity: just as Lorentz invariant quantities are independent of the choice of inertial frame in special relativity,  the probabilities derived from the physical observable $\hat A_0$ are the same in both QRFs $\A$ and $\B$. A more complete formulation of special relativity in the perspective neutral approach can be found in~\cite{delahamette2021perspectiveneutralapproachquantumframe}.

\subsubsection{A key difference with alternative proposals: the physical Hilbert space}

The space of global states in the perspective neutral approach is more restricted than in the other approaches: global state vectors live in $\ch_0$, whereas in the other approaches they are block diagonal density operators on $\ch_0$ and $\ch_1$ (equivalently, density operators invariant under the global action). 

Indeed, $\cb(\ch_0)$ is the image of $\Pi_0(\cdot) \Pi_0$ which is strictly contained in $\cb(\ch_0) \oplus \cb(\ch_1)$, the image of $\cg$,  and since both share the same domain it follows that $\Pi_0 (\cdot) \Pi_0$ `removes more information' than $\cg$~\cite{hoehn2023quantumframerelativitysubsystems}.   This information is the `which global charge' (here, labelling $\ch_0$ or $\ch_1$), which is an extra degree of freedom present in the extra particle and operational approaches. This degree of freedom is invariant under the group of interest. In our example, $\Pi_0$ and $\Pi_1$, the projectors onto $\ch_0$ and $\ch_1$, respectively, commute with the global $\mathbb{Z}_2$ action. Measuring these projectors might require an implicit frame for the conjugate quantity of the symmetry group. In the case of global translation symmetry the conjugate quantity is center of mass momentum. The physical Hilbert space in this case is motivated by invariance under time dependent global translations~\cite{Vanrietvelde2020changeof}, a stronger constraint than just global translation invariance. 
A { separate} argument is given in~\cite[II, B,3]{hoehn2023quantumframerelativitysubsystems} in favour of removing this degree of freedom, using the coherent twirl, in contrast to approaches which keep it.

\subsection{Extra particle approach}

From our main argument, it is clear that knowing the state of $\B$ and $\C$, relative to Alice's QRF $\A$ is not sufficient for her to determine unambiguously the state of $ \A$ and $ \C$ relative to Bob's QRF $\B$. She needs more information. Specifically, Alice needs to know whether the global state lies in $\mathcal{H}^0$ (trivial representation) or $\mathcal{H}^1$ (sign representation).

The extra particle formalism \cite{castroruiz2023relative} captures the degrees of freedom that are measureable in the absence of a reference frame for a group $G$ -- $\mathbb{Z}_2$ in this case. This singles out $G$-invariant operators, including projectors onto distinct charge sectors. According to this formalism, the `which irrep.' information is accessible in principle by the agents. This is the most natural interpretation of the extra-particle formalism. Note that this extra information is $\mathbb{Z}_2$-invariant, as it is encoded in the subsystem formed by the global charge of $\sf{A}$, $\sf{B}$ and $\sf{C}$. This subsystem is called the extra particle (hence the name of the formalism). Being $\mathbb{Z}_2$-invariant, the `which irrep.' information is independent of any external reference frame for $\mathbb{Z}_2$, which, by assumption, Alice has no access to. Note that lacking access to a reference frame for $\mathbb{Z}_2$ does not imply lacking access to an implicit ($\mathbb{Z}_2$-invariant) reference frame for charge, with respect to which the charge of $\sf{ABC}$ may be measured. Recent works have studied the subtleties that may arise when measuring total charge, formulating thought experiments where different parties collaborate to this end \cite{delahamette2025quantum, doat2025can}.

Now, note that, under stronger symmetry constraints, such `which irrep.' variables may no longer be accessible to $\A$ and $\ B$. In this case, these variables take on a different interpretation in the extra particle approach: rather than representing accessible degrees of freedom, they serve as conventional values needed to uniquely specify the transformation between different QRFs. In our case, the extra-particle variables, when inaccessible, can be understood as fixing a conventional value for the total momentum of systems $A$, $B$, and $C$. As our example illustrates, these conventional values are important even when they are not physically measurable — Alice can determine the state of $A$ and $S$ `as seen' from Bob only if she knows which convention he is using.

The extra particle formalism can be derived from standard quantum theory by demanding consistency with the hypothetical existence of external reference frames, and by assuming that relative subsystems with respect to a QRF are given by the algebra of relative operators defined in Appendix \ref{sec:observable_based}.

In this formalism, Alice  divides `the world' into operationally defined relative subsystems: the subsystem of $\sf B$ relative to $\sf A$, the subsystem of $\sf C$ relative to $\sf A$ and the `extra particle', denoted by $\overline{\sf{BC}}$. A QRF transformation means a change from Alice's preferred subsystem decomposition to Bob's preferred subsystem decomposition\footnote{The extra particle formalism is thus consistent with the quantum relativity of subsystems  which occurs in the perspective neutral framework~\cite{hoehn2023quantumframerelativitysubsystems}.}. As our main argument shows, the extra particle subsystem is key for the reversibility of QRF transformations in this formalism. 

For a detailed exposition of the extra particle framework, a we recommend the interested reader to consult Ref. \cite{castroruiz2023relative}. There, arguments are given in favour of the incoherent twirl as the appropriate tool to describe physics when agents lack access to an external reference frame (including a discussion of the argument presented in \cite{hoehn2023quantumframerelativitysubsystems}).  In this section, we limit ourselves to describing how Alice can apply the formalism and reconstruct the state of $\A$ and $\C$ relative to $\B$ from her state on $\sf{BC}$ and $\overline{\sf{BC}}$ relative to $\sf A$. The procedure can be explained as follows. First, Alice determines whether the extra particle state (i.e. the state of the subsystem $\overline{\sf{BC}}$) lives in $\ch_0$ or $\ch_1$. If the answer is $\ch_0$ (trivial representation), she applies the unitary 
\begin{equation}
U_+ = \ketbra{0}{0}^\A_\B \otimes \I^\A_\C + \ketbra{1}{1}^\A_\B \otimes X^\A_\C  .
\end{equation}
If the answer is $\ch_0$ (sign representation), she applies 
\begin{equation}
U_- = \ketbra{0}{0}^\A_\B \otimes \I^\A_\C - \ketbra{1}{1}^\A_\B \otimes X^\A_\C .
\end{equation}
Therefore, the full quantum reference frame transformation in the extra particle approach is
\begin{equation}
S_{\sf A \rightarrow \sf B} = \ketbra{0}{0}^\A_{\overline{\sf{BS}}} \otimes U_+ + \ketbra{1}{1}^A_{\overline{\sf{BS}}} \otimes U_-.
\end{equation}
By construction, this procedure maps any state $\ket{\pm}^\A_{\overline{\sf{BC}}} \otimes \ket{\chi}^\A_{\sf{BC}}$ to the corresponding state $\ket{\pm}^\B_{\overline{\sf{AC}}} \otimes \ket{\chi}^\B_{\sf{AC}}$, where $\ket{\chi}^\A_{\sf{BC}}$ is obtained by applying the map $\calr_{\bra 0_\A}$ to a global pure state as explained before \Cref{zerowrtA}. After applying the QRF transformation, the state of $\sf A$ and $\sf C$ with respect to $\sf B$ is simply obtained by tracing out the subsystem $\overline{\sf{AC}}$. When the global state is $\ket{\psi^0}$ ($\ket{\psi^1}$), the resulting state is given by $\ket{\psi^0} = (\ket{01}-\ket{10})/\sqrt{2}$ ($\ket{\psi^1} = (\ket{01}+\ket{10})/\sqrt{2}$), in agreement with Eqs.~(\ref{statefrombob0}, \ref{statefrombob1}). This change of QRF agrees with the perspective neutral approach on the $\ch_0$ subspace and extends it to the $\ch_1$ subspace.

Finally, we observe that the state of $\B\C$ relative to $\A$ can equivalently be obtained from the global state by a `disentangling transformation', defined as 
\begin{equation}\label{eq:disentangler}
T^\A_{\B \C} = \ketbra{0}{0}_\A \otimes \I_\B \otimes\I_\C + \ketbra{1}{1}_\A \otimes X_\B \otimes X_\C. 
\end{equation}
This transformation can be interpreted as a change from the subsystem decomposition relative to any external observer to the subsystem decomposition that is naturally adapted to Alice's QRF.

\subsection{Operational approach}

A final resolution is to take the view that states that cannot be distinguished by Alice or Bob are viewed, by them, as equivalent. Concomitantly, it appears as though information is lost in comparison to the other approaches. For instance, the charge of the global state is not accessible by means of observables relative to $\sf A$. Alice and Bob must work only with what they can measure, and all information accessible only to an `all seeing eye' must be discarded at the subsystem level.

For instance, suppose the state relative to Alice is $\ket{\phi}_{\B\C}^\A = \frac{1}{\sqrt{2}} (\ket{01} - \ket{11})$ (see \cite[Subsection 3.3]{carette2025operational}  for an operational justification of the notion of `relative state')  which is compatible with the global state lying in $\ch_0$ or $\ch_1$. As we have seen, the distinct global states give distinct states relative to Bob, and therefore it looks like there is no viable map effecting the frame transformation. However, the following stipulation can be made: states relative to Bob that cannot be distinguished by Alice's frame are identified. Mathematically, this can be done by considering observables of the form $\ket{i}\bra{i}_{\A} \otimes T_\C$ (where $i=0,1$ and $T_\C$ is an arbitrary observable of Charlie) and their linear combinations (these are the \emph{framed} observables - see \cite[Subsection 4.2]{carette2025operational}; these reflect the idea that for a given experiment, a particular observable is singled out for performing the function of a reference frame), or the stronger constraint that only observables of Charlie relative to Alice are considered. Either way, the `situation' relative to Bob is described by Alice only up to what can be distinguished relative to Alice's frame. 

Consider again the states 
\begin{align}
    \ket{\psi^0}_{\A\C}^\B = \frac{1}{\sqrt{2}} (\ket{01} - \ket{10}), \\
    \ket{\psi^1}_{\A\C}^\B = \frac{1}{\sqrt{2}} (\ket{01} + \ket{10}),\
\end{align}
relative to Bob. A short calculation reveals that for any observable $T_\C$ of Charlie's system, 
\begin{equation}
    \bra{\psi^0}^\B_{\A\C}(\ket{i}\bra{i}_{\sf A} \otimes T_\C)\ket{\psi^0}^\B_{\A\C} = \bra{\psi^1}^\B_{\A\C}(\ket{i}\bra{i}_{\sf A} \otimes T_\C)\ket{\psi^1}^\B_{\A\C},
\end{equation}
for $i=0$ or $1$. This conclusion holds for linear or convex combinations, and therefore in particular holds also for observables of $\A$ relative to $\C$. Therefore the distinct states relative to Bob's frame cannot be distinguished by Alice. These may be thus identified, along with all other such equivalent states, to form an equivalence class. 

Since Alice has no reason to favour either $\ket{\psi^0}$ or $\ket{\psi^1}$ as giving rise to Bob's relative state, a reasonable approach may be for her to choose to give Bob's two possible assignments $\ket{\psi^0}_{\A\C}^\B$ or $\ket{\psi^1}_{\A\C}^\B$ equal probability, reflecting her complete ignorance, resulting in the 
 mixed state: 
 
\begin{align}
   \frac{1}{2} \left(\ketbra{\psi^0}{\psi^0}^\B_{\A\C} + \ketbra{\psi^1}{\psi^1}^\B_{\A\C}\right) = \frac{1}{2} \left(\ketbra{01}{01}_{\A\C} + \ketbra{10}{10}_{\A\C} \right) = \bar \rho^\B_{\A\C}, \label{eq:op_state_equiv}
\end{align}
as Bob's relative state. This supposition can be justified by noting that the state above lies in the same equivalence class that 
$\ket{\psi^0}_{\A\C}^\B$ and $\ket{\psi^1}_{\A\C}^\B$ are mapped to under the equivalence relation delineated above, as can be calculated by checking $\Tr[\bar \rho^\B_{\A\C} \ket{i}\bra{i} \otimes T_\C]$.

A more complete treatment of this scenario in the operational approach is given in the Appendices. This includes a derivation of the map from Alice's relative state $\rho_{\B\C}^\A$ to the state $\bar \rho_{\A\C}^\B$ of~\Cref{eq:op_state_equiv}. Moreover it is shown that the framed observables are exactly those which are accessible to both Alice and Bob, and hence are exactly the observables that Alice would use to make inferences about Bob's relative state. 

At the level of quantum states (pure or mixed) the frame change as described above is not invertible and, a fortiori, not even a map, since Alice's relative state corresponds to more than one state relative to Bob. However, by forming the described equivalence classes, the frame change is a well defined map, whose output is not the set of quantum states but the convex set of equivalence classes of quantum states. If such an equivalence class is also imposed on Alice, this map is actually invertible at the level of convex sets, but is clearly not in any sense unitary (see~\Cref{eq:op_QRF_change}).

\begin{table}
    \centering
    \begin{tabular}{|c|c|c|c|c|} 
    \hline
         Approach & Physical states & Change of QRF & Can $\A$ distinguish all global physical states?   \\
         \hline 
        Perspective-Neutral & $\Pi_0 \ket \psi = \ket \psi \in \ch_0$ & Invertible & Yes  \\
        Operational & $\cg(\rho) = \rho \in \cb(\ch_0) \oplus \cb(\ch_1)$ & Non-invertible  & No  \\ %& Inferential \\
        Extra-particle & $\cg(\rho) = \rho \in \cb(\ch_0) \oplus \cb(\ch_1)$ & Invertible &  Yes\\ %& Inferential
         \hline 
    \end{tabular}
    \caption{Summary of the three different approaches and their characteristics.}
    \label{tab:summary}
\end{table}

\begin{figure}
    \centering
    \includegraphics[width=
    \linewidth]{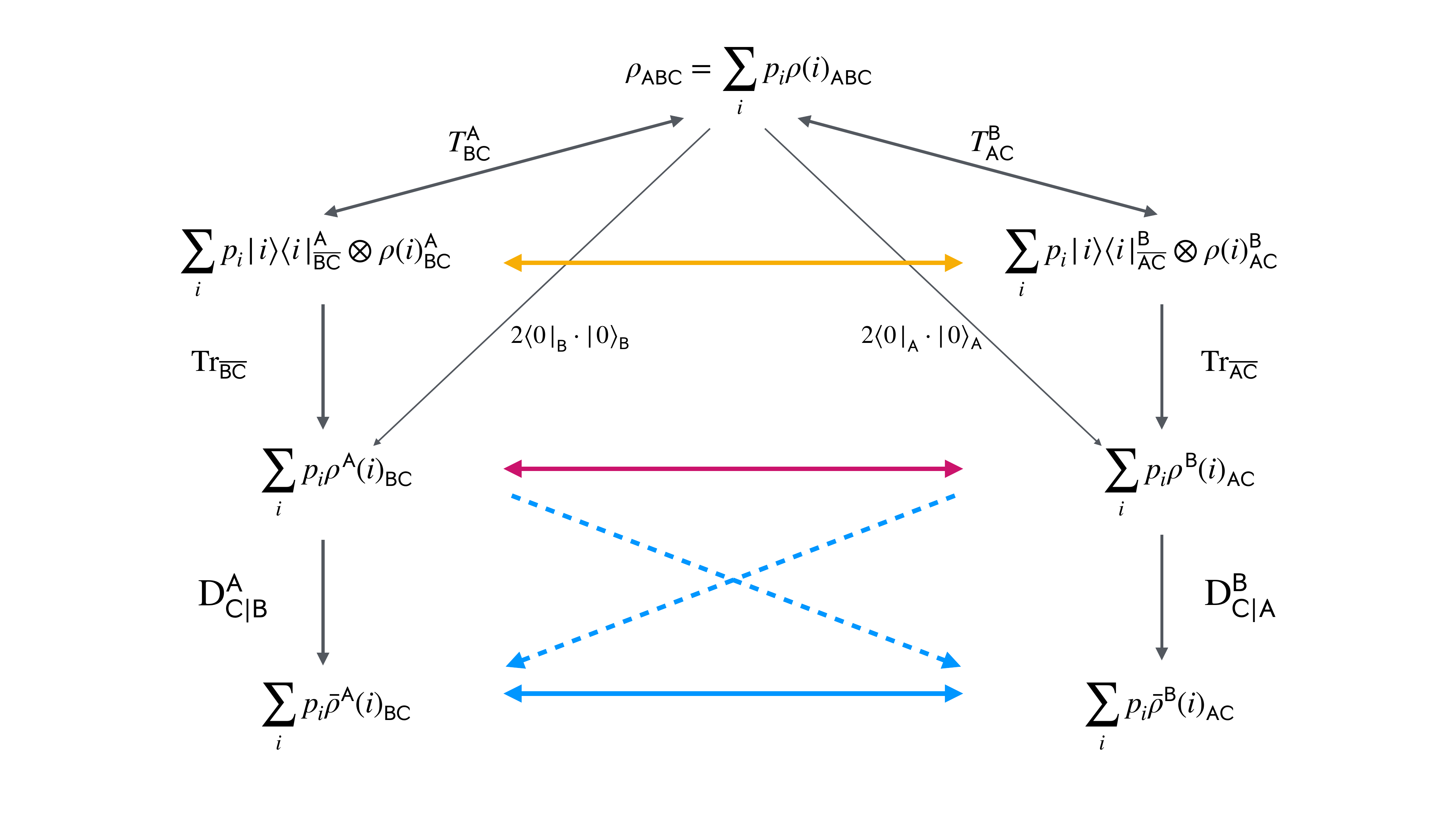}
    \caption{This diagram illustrates the changes of QRF in different approaches. Yellow: Extra particle approach. The map $T^\A_{\B\C}$ is defined in Eq.~(\ref{eq:disentangler}) and an analogous definition holds for $T^\B_{\A\C}$. Purple: perspective neutral approach (when restricted to the sector $i = 0$) . When $\rho_{\A\B\C}$ has support on multiple sectors then the purple arrow does not denote a well defined map. Blue: operational approach. The dashed lines represent non-invertible maps, while the solid line defines an invertible map. $\mathrm{D}^{\sf A}_{\sf{C|B}}$ denotes a non-selective measurement in the $\ket{0}^\A_\B, \ \ket{1}^\A_\B$ basis and  $\mathrm{D}^{\sf B}_{\sf{C|A}}$ denotes a non-selective measurement in the $\ket{0}^\B_\A, \ \ket{1}^\B_\A$ basis (see Appendix \ref{appframed}).}
    \label{fig:outline_of_main_arg2}
\end{figure}

\subsection{Comparison between the approaches}

\subsubsection{Global states}

The three approaches differ in the allowed global states: the perspective neutral approach only considers global states which lie in $\ch_0$ whilst the extra-particle and operational approach allow invariant density operators (which are in $\cb(\ch_0) \oplus \cb(\ch_1)$ in our example). The perspective neutral approach imposes invariance at the level of the state vector: $U(g) \ket \psi = \ket \psi$, whilst the other two approaches impose invariance at the level of the density operator: $U(g) \rho U^\dagger(g) = \rho$. 

The different requirements imposed on the global states point to important physical differences between the frameworks: in the presence of constraints the perspective neutral more naturally applies, whilst in the absence of constraints the operational and extra-particle more naturally apply.

\subsubsection{Accessible observables}

Another key difference between the frameworks is the accessible observables: whilst the extra-particle and operational frameworks have the same global states, they consider different accessible observables. In the operational framework the accessible observables to QRF A are just the observables $\yen^\A_{\B\C}(T_{\B\C})$ for $T_{\B\C} \in \cb(\ch_{\B\C})$, which form a strict subalgebra of the invariant observables $\cb(\ch_{\A\B\C})^G$. In the extra-particle approach the QRF A also has access to the extra-particle $\overline{\B\C}$ which together with the observables of the form  $\yen^\A_{\B\C}(T_{\B\C})$  span the full invariant algebra. This highlights a crucial physical distinction between the notion of reference frame in both approaches: in the operational the QRF's $\A$, $\B$ and $\C$ are associated to different subalgebras of $\cb(\ch_\A \otimes \ch_\B \otimes \ch_\C)^G$ whereas in the extra-particle approach different reference frames correspond to different decompositions of the same global algebra  $\cb(\ch_\A \otimes \ch_\B \otimes \ch_\C)^G$. This distinction is leads to a crucial conceptual difference between the two frameworks, namely the invertibility (or not) of the change of quantum reference frame.

In the perspective neutral approach the observables accessible to QRF $\A$ initially appear to be those of the form $\yen^\A_{\B\C}(T_{\B\C})$  as in the operational approach. However since the global observables are $\cb(\ch_0) \subsetneq \cb(\ch_\A \otimes \ch_\B \otimes \ch_\C)^G$ one must restrict the observables $\yen^\A_{\B\C}(T_{\B\C})$  to $\ch_0$. With this restriction one can show that the observables associated to $\A, \B$ and $\C$ are all isomorphic to $\cb(\ch_0)$, and hence different reference frames correspond to different decompositions of the global algebra $\cb(\ch_0)$. The perspective neutral approach and extra-particle approach share an important conceptual understanding of QRFs as corresponding to different decompositions of the global algebra, they just differ on what the global algebra is.

\section{Discussion}

\subsection{Purely perspectival approach}

The `purely perspectival' approaches to QRFs of~\cite{Giacomini_2019,delaHamette2020quantumreference} derive changes of QRF of the form $\ket{\psi}^{\A}_{\B\C} \mapsto \ket{\psi}^{\B}_{\A\C}$   without explicit appeal to a global state. Such approaches could be understood as having an implicit global state, and indeed some of the changes of QRFs presented in those works have been derived within the perspective neutral framework~\cite{Vanrietvelde2020changeof,delahamette2021perspectiveneutralapproachquantumframe}. Our example illustrates that if a global state is implicitly assumed then the change of QRF in the purely  perspectival approach does in fact implicitly make an appeal to some feature of the global state, namely its global charge, in order to justify which of the compatible states relative to $\B$ is mapped to from the state relative to $\A$. 

Depending on the physical situation, the global charge might be accessible or inaccessible to Alice and Bob (being invariant, the lack of access to an external frame need  not preclude Alice and Bob from measuring it).  In the case where the charge is inaccessible, Alice and Bob can still assign a conventional value to it, as if it were measured `externally'. This value then fixes the QRF transformation that they use, since, as shown in the extra-particle approach the change of QRF depends on the global charge. Importantly, Alice and Bob have to assign the same value to the global charge by convention. 

Alternatively, a more radical interpretation of the purely perspectival approach is to reject the presence (even implicit) of a global state. However, in~\cite{delaHamette2020quantumreference} this requires appealing to a `principle of coherent change of reference system' which involves first defining the change of QRF for a classical basis $\{\ket{\psi_i}_{\A\B}^\C\}_i$: $\ket{\psi_i}_{\A\B}^\C \mapsto \ket{\psi_i}_{\B\C}^\A $ and then extending linearly to $\sum_i \alpha_i \ket{\psi_i}_{\A\B}^\C \mapsto \sum_i \alpha_i\ket{\psi_1}_{\B\C}^\A$. Here the relative phases are arbitrarily set to 0, but any generic transformation $\sum_i \alpha_i \ket{\psi_i}_{\A\B}^\C \mapsto \sum_i \alpha_i e^{i \theta_i} \ket{\psi_i}_{\B\C}^\A$ would constitute a valid extension of the transformation on the classical basis.

As has been shown in~\cite{Hoehn_2022}, whilst the approach of~\cite{delaHamette2020quantumreference} can be recovered from the perspective neutral approach for kinematics, it disagrees for the possible dynamics. Further details on the differences between the perspective neutral and purely perspectival approaches can be found in~\cite[Section 10]{delahamette2021perspectiveneutralapproachquantumframe}.

\subsection{The limitations of the example}\label{sec:limitations}

For pedagogical reasons we have focused on the simplest non-trivial example: the group $\Zl_2$ is compact and abelian,   the systems $\Cl^2$  are finite dimensional and the reference frames are ideal (a reference frame for a finite group $G$ is ideal when the coherent state system (i.e., the orbit of a vector under a unitary representation) $\{\ket{g}|g \in G\}$ is such that $\braket{g|h} = \delta_{g,h}$ for all $g, h \in G$).  An example of a non-ideal frame for a qubit is given by the coherent state system $\{\ket 0, \ket +, \ket 1, \ket -\}$ which transforms under the representation of $\Zl_4$ generated by the rotation $e^{i Y \frac{\pi}{4}}$. The associated covariant POVM is given by $\{\frac{\ketbra{0}{0}}{2}, \frac{\ketbra{+}{+}}{2},\frac{\ketbra{1}{1}}{2},\frac{\ketbra{-}{-}}{2}\}$.

This entails that features that further distinguish the different approaches may not be apparent in this example. However, this example does highlight several key differences between the approaches, which are present in other more general scenarios, such as the invertibility (or not) of the change of reference frame map and the use of the coherent or incoherent twirl to define the global states.

\subsection{Related work}

Existing comparisons between different approaches to quantum reference frames include the following:
\begin{itemize}
    \item In~\cite{Krumm2021quantumreference}, the quantum information approach of~\cite{bartlett2007reference} is contrasted to the structural approach, which includes the purely perspectival and perspective neutral approaches. A communication protocol is introduced which highlights the difference between the two approaches: the quantum information approach deals with the transmission of quantum systems whilst the structural approach concerns agents coordinating their description of quantum states.
    \item In~\cite{Hoehn_2022}, the perspective neutral approach and the `alignable' approach (which is the specific purely perspectival approach of~\cite{delaHamette2020quantumreference}) are contrasted and shown to be dynamically inequivalent, despite kinematical equivalence (in the case of ideal reference frames).
    \item A comparison between the operational approach and the purely perspectival approach is given in~\cite[Section 4.4]{carette2025operational} highlighting the fact that, as in our example, the coherence of the  change of quantum reference frame of~\cite{Giacomini_2019,delaHamette2020quantumreference} is not  accessible to the agents in the operational approach. An analysis is undertaken showing that the operational frame change is not inconsistent with either the perpective neutral or purely perspectival approach on the common regime in which they are defined.
    \item In~\cite{castroruiz2023relative}, the extra-particle approach is shown to recover the purely perspectival changes of quantum reference frame and the perspective neutral changes of quantum reference frame for ideal frames (as defined in~\Cref{sec:limitations}). 
    \item In~\cite{devuyst2025relation}, the relation between the perspective neutral, algebraic and effective approaches to quantum reference frames is explored, showing that the three are equivalent for ideal reference frames.  Unlike the perspective neutral approach which is based on Hilbert space, both the algebraic and the effective  approaches are built on a quantum phase space, consisting of expectation values of operators (including their higher-order moments). The effective approach is more restrictive than the algebraic approach since it implements a semi-classical approximation (at some order of $\hbar$).
    \item In~\cite{delahamette2021perspectiveneutralapproachquantumframe}, the changes of quantum reference frame for ideal frames in the purely perspectival case is recovered within the perspective neutral framework, and moreover it is shown that the two formalisms disagree for non-ideal frames.
\end{itemize}

Finally, we note that while our work, and the other work mentioned here, highlight the differences between various approaches to quantum references, there also exist cases where different approaches can co-exist within the same setup. The perspective neutral and quantum information approaches have been shown to co-exist within the context of quantum error correction~\cite{carrozza2024correspondence} and (also with the operational approach) within lattice gauge theories~\cite{araujoregado2025relational}.

\section{Conclusion}

By using a simple example we have shown the key differences between three recent approaches to quantum reference frames: the perspective neutral, extra-particle  and operational  approaches. These approaches disagree on the global states, with the perspective neutral approach adopting states on the charge zero sector  whilst the extra-particle and operational approaches allow invariant states with support outside the zero charge sector.  In both the PN and extra-particle approach the perspectives of different QRFs induce different decompositions of the global algebra, whereas in the operational approach different (ideal) QRFs correspond to different sub-algebras of the global algebra. Hence in the perspective neutral and extra-particle approaches, all perspectives are equivalent to the global one, whereas in the operational approach they are not. Potential directions for future work include extending this example to from $\Zl_2$ to non-abelian and continuous groups and from ideal to non-ideal frames in order to determine whether further differences between the frameworks emerge.  

\section*{Acknowledgments}

T. G. acknowledges helpful discussions with Jan G\l{}owacki, whom we also thank for some valuable feedback on an earlier draft. E. C-R. acknowledges helpful discussions with Ognyan Oreshkov and Carlo Rovelli.  E. C-R. and T. G. acknowledge helpful discussions with Anne-Catherine de la Hamette, Guilhem Doat, Viktoria Kabel and Augustin Vanrievelde. We also thank Anthi Voulgari Revof for helpful feedback. This research was funded in part by the Austrian Science Fund (FWF) 10.55776/PAT2839723, via the Austrian Science Fund (FWF) project P 33730-N, and via the Austrian Science Fund (FWF) COE 1 Quantum Science Austria.

\bibliographystyle{unsrt}
\bibliography{references}

\appendix

\section{Observable based description}
\label{sec:observable_based}

In the initial presentation of the example there was little mention of the observables. We present an equivalent formulation of the example which makes explicit reference to the observables that the agents can measure. 

As discussed in e.g. \cite{loveridge2012quantum,glowacki2024quantum,miyadera2016approximating,loveridge2018symmetry,carette2025operational}, the observables relative to Alice are given by the following map:
\begin{align}
    \yen_{\B\C}^\A:& \cb(\ch_\B \otimes \ch_\C) \to \cb(\ch_\A \otimes \ch_\B \otimes \ch_\C)^{G} \subset \cb(\ch_\A \otimes \ch_\B \otimes \ch_\C) , \\
    \yen_{\B\C}^\A:& T_{\B\C} \mapsto \ketbra{0}{0}_\A \otimes T_{\B\C} + \ketbra{1}{1}_\A \otimes (X_\B \otimes X_\C)T_{\B\C}(X_\B \otimes X_\C)\label{eq:relativeobservable}.
\end{align}

The interpretation of this map is that it relativises an
observable $T_{\B\C}$, relative to the $\mathbb{Z}_2$-covariant reference frame observable (projection-valued measure) $Z_\A = \{ \ketbra{0}{0}_\A,\ketbra{1}{1}_\A\}$. 

Importantly, the relativized observable $\yen_{\B\C}^\A(T_{\B\C})$ is invariant under the global action, and is therefore defined and understood independently of any external reference frame. The relativisation map $\yen_{\A\C}^{\B}$ can be obtained by a suitable permutation of the subsystem labels. 

Straightforward computation shows that:
\begin{align}
    \bra{\psi^0} \yen_{\B\C}^{\A}(T_{\B\C}) \ket{\psi^0}= \bra{\psi^0}^\A_{\B\C} T_{\B\C} \ket{\psi^0}^\A_{\B\C} , \\
     \bra{\psi^1} \yen_{\B\C}^{\A}(T_{\B\C}) \ket{\psi^1}= \bra{\psi^1}^\A_{\B\C} T_{\B\C} \ket{\psi^1}^\A_{\B\C} .
\end{align}
Since this holds for all $T_{\B\C}$ and since $ \ket{\psi^0}^\A_{\B\C}  =  \ket{\psi^1}^\A_{\B\C} $ it follows that observables relative to Alice's reference frame cannot distinguish the global state $\ket{\psi^0}$ from $\ket{\psi^1}$.

Similarly:
\begin{align}
    \bra{\psi^0} \yen_{\A\C}^{\B}(T_{\A\C}) \ket{\psi^0}= \bra{\psi^0}^\B_{\A\C} T_{\A\C} \ket{\psi^0}^\B_{\A\C} , \\
     \bra{\psi^1} \yen_{\A\C}^{\B}(T_{\A\C}) \ket{\psi^1}= \bra{\psi^1}^\B_{\A\C} T_{\A\C} \ket{\psi^1}^\B_{\A\C}  .
\end{align}
Since $\ket{\psi^0}^\B_{\A\C}  \neq  \ket{\psi^1}^\B_{\A\C}$ there exists some $T_{\A\C}$ which allow Bob, with his relative observables, to distinguish the  global state  $\ket{\psi^0}$ from $\ket{\psi^1}$. 

The adjoint (\emph{pre-dual}) map $(\yen^\A_{\B\C})_*$, defined by  
\begin{align}\label{eq:yenstar}
    \Tr(\rho_{\A\B\C} \yen_{\B\C}^{\A}(T_{\B\C})) = \Tr((\yen^\A_{\B\C})_*(\rho_{\A\B\C}) T_{\B\C})
\end{align}
and holding for all $T_{\B\C}$, takes the form 
 $(\yen^\A_{\B\C})_*: \rho_{\A\B\C} \mapsto \rho_{\B\C}^{\A} = 2 \bra{0}_\A \rho_{\A\B\C} \ket{0}_\A$ on invariant states, which is the map $\calr_{\bra{0}_\A}$ extended to density operators: $\calr_{\bra{0}_\A} \rho_{\A\B\C} \calr_{\bra 0_\A}^\dagger$.
Eq. \eqref{eq:yenstar} shows that evaluating relative observables  on global states is equivalent to evaluating local observables on the relative states:
\begin{align}
     \Tr(\rho_{\A\B\C} \yen_{\B\C}^{\A}(T_{\B\C}))  = \Tr(\rho_{\B\C}^{\A} T_{\B\C})
\end{align}

The unitary $X_\A \otimes X_\B \otimes X_\C$ is also a Hermitian operator, with eigenvalues $\pm 1$. It is invariant under the action of the symmetry group, but not a relative observable in the sense defined above. We observe that:
\begin{align}
    \bra{\psi^0}X_\A \otimes X_\B \otimes X_\C \ket{\psi^0} \neq  \bra{\psi^1}X_\A \otimes X_\B \otimes X_\C \ket{\psi^1},
\end{align}
hence access to this observable would allow Alice to distinguish between the two global states (but this is prohibited for internal observers in the operational approach). This encodes the information contained in the extra-particle approach of~\cite{castroruiz2023relative}.

\section{Framed observables and invertible changes of quantum reference frame in the operational approach}\label{appframed}

Given a reference frame $\B$ with covariant PVM $\{\ketbra{i}{i}_\B\}_{i = 0,1}$ framed observables on $\ch_\B \otimes \ch_\C$ are of the form $\sum_i \ketbra{i}{i}_\B \otimes C_i$~\cite{carette2025operational}. Note that in this case of an ideal frame, the framed observables form an algebra. Given two global states $\rho_{\A\B\C}$ and $\rho^{\prime}_{\A\B\C}$ we have the following:
\begin{align}
    \Tr(\rho_{\A\B\C} \yen_{\B\C}^{\A}(T_{\B\C})) = \Tr(\rho^{\prime}_{\A\B\C} \yen_{\B\C}^{\A}(T_{\B\C})) \ \forall T_{\B\C} \in \cb(\ch_\B \otimes \ch_\C) \implies \rho_{\B\C}^\A = \rho^{\prime \A}_{\B\C} , \\
    \exists \  T_{\B\C} \in \cb(\ch_\B \otimes \ch_\C)  \ {\textrm{s.t.}} \ \Tr(\rho_{\A\B\C} \yen_{\B\C}^{\A}(T_{\B\C})) \neq \Tr(\rho^{\prime}_{\A\B\C} \yen_{\B\C}^{\A}(T_{\B\C})) \ \implies \rho_{\B\C}^\A \neq \rho^{\prime \A}_{\B\C}
\end{align}

Hence the equivalence classes of global states that can be distinguished by Alice's relative observables $\yen_{\B\C}^{\A}(T_{\B\C})$ are in one to one correspondence with the relative states $\rho_{\B\C}^\A$. See \cite{carette2025operational} subsec. 3.3.2 for a construction of relative state spaces, and Prop. 3.11 for the general statement of the relevant isomorphism.

Given a state $\rho_{\B\C}^\A$, the operational approach then defines a further equivalence class, given by states that cannot be distinguished by framed observables relative to $\B$. Two states $\rho_{\B\C}^{\A}$ and $\rho^{\prime \A}_{\B\C}$ are equivalent if they agree on all $\B$-framed observables:
\begin{align}
    \rho_{\B\C}^{\A} \sim \rho^{\prime \A}_{\B\C} \iff \Tr(\rho_{\B\C}^{\A} (\sum_i \ketbra{i}{i}_\B \otimes C_i) )=\Tr(\rho_{\B\C}^{\prime \A} (\sum_i \ketbra{i}{i}_\B \otimes C_i)), \forall C_i \in \cb(\ch_\C).
\end{align}
Note that for every $\rho_{\B\C}$, the state $\bar \rho_{\B\C}:=\sum_k \ketbra{k}{k}_\B \otimes \bra{k}_\B \rho_{\B\C} \ket k_\B$ gives the same expectation values for all framed observables $\sum_i \ketbra{i}{i}_\B \otimes C_i$:
\begin{align}
     \Tr(\bar \rho_{\B\C} (\sum_i \ketbra{i}{i}_\B \otimes C_i))=& \Tr((\sum_k \ketbra{k}{k}_\B \otimes \bra{k}_\B \rho_{\B\C} \ket k_\B)(\sum_i \ketbra{i}{i}_\B \otimes C_i)) \\
    =& \sum_{ijkl} \bra{jl}_{\B\C} (\sum_k \ketbra{k}{k}_\B \otimes \bra{k}_\B \rho_{\B\C} \ket k_\B)(\sum_i \ketbra{i}{i}_\B \otimes C_i) \ket{jl}_{\B\C} \\
    =& \sum_{jl} \bra{j}_\B \bra{l}_\C \rho_{\B\C} \otimes C_j  \ket{j}_\B \ket{l}_\C\\
    =& \sum_{jl} \bra{j}_\B \bra{l}_\C \rho_{\B\C}( \sum_i \ketbra{i}{i}_\B  \otimes C_i)  \ket{j}_\B \ket{l}_\C\\
    =& \Tr(\rho_{\B\C} (\sum_i \ketbra{i}{i}_\B \otimes C_i)).
\end{align}
We check that the map $\rho_{\B\C} \mapsto \bar \rho_{\B\C}$ is trace preserving:
\begin{align}
    \Tr(\bar \rho_{\B\C}) = \sum_{ij} \bra{ij}_{\B\C} (\sum_k \ketbra{k}{k}_\B \otimes \bra{k}_\B \rho_{\B\C} \ket k_\B   \ket{ij}_{\B\C} = \sum_{ij} \bra{ij}_{\B\C}\rho_{\B\C}  \ket{ij}_{\B\C} = \Tr(\rho_{\B\C}) .
\end{align}

Therefore we can define a frame reduction map  $\mathrm{D}^\A_{\C|\B}: \rho_{\B\C} \mapsto \bar \rho_{\B\C} = \sum_k \ketbra{k}{k}_\B \otimes \bra{k}_\B \rho_{\B\C} \ket k_\B$ (where we retain the superscript $\A$ as a reminder that $\rho_{\B\C}$ and hence $\bar \rho_{\B\C}$ will typically be obtained from a global state $\rho_{\A\B\C}$). This map is  linear and idempotent: it is a projector that maps states with support on $\cb(\ch_\B \otimes \ch_\C)$ to states with support on the subalgebra of framed observables relative to $\B$. 

Given an invariant (global) state $\rho_{\A\B\C}$ we obtain:
\begin{align}
    \bar \rho_{\B\C}^\A =  \mathrm{D}_{\C|\B}^\A(\calr_{\bra 0_\A} \rho_{\A\B\C} \calr_{\bra 0_\A}^\dagger) =  2  \sum_k \ketbra{k}{k}_\B \otimes \bra 0_\A \bra{k}_\B \rho_{\A\B\C} \ket 0_\A \ket k_\B, \\
    \bar \rho_{\A\C}^\B = \mathrm{D}_{\C|\A}^\B(\calr_{\bra 0_\B} \rho_{\A\B\C} \calr_{\bra 0_\B}^\dagger) = 2  \sum_k \ketbra{k}{k}_\A \otimes \bra 0_\B \bra{k}_\A \rho_{\A\B\C} \ket 0_\B \ket k_\A.
\end{align}

We define the map 
\begin{align}
    Q_{\C|\B^\A \to \C|\A^\B}(\bar \rho_{\B\C|\A} ) = \sum_i \bra{i}_\B \otimes X^i_\C  (\bar \rho_{\B\C|\A} ) X^i_\C \ket{i}_\B \otimes \ketbra{i}{i}_\A. \label{eq:op_QRF_change}
\end{align}

For an invariant state $\rho_{\A\B\C}$ and $\bar \rho_{\B\C|\A} = \mathrm{D}_{\C|\B}^\A(\calr_{\bra 0_\A} \rho_{\A\B\C} \calr_{\bra 0_\A}^\dagger)$ we obtain:
\begin{align}
     Q_{\C|\B^\A \to \C|\A^\B}(\bar \rho_{\B\C|\A} )  =& \sum_i \bra{i}_\B \otimes X^i_\C  (\bar \rho_{\B\C|\A} ) X^i_\C \ket{i}_\B \otimes \ketbra{i}{i}_\A = 2 \sum_i \ketbra{i}{i}_\A \otimes \bra 0_\A \bra{i}_\B X^i_\C \rho_{\A\B\C} X^i_\C \ket 0_\A \ket i_\B \\
    =& 2  \sum_i \ketbra{i}{i}_\A \otimes \bra 0_\B \bra{i}_\A \rho_{\A\B\C} \ket 0_\B \ket i_\A = \bar \rho_{\A\C|\B} ,
\end{align}
where we have used the identity $X^i_\C \rho_{\A\B\C} X^i_\C = X^i_\A \otimes X^i_\B \rho_{\A\B\C} X^i_\A \otimes X^i_\B$ and $\bar \rho_{\A\C|\B} = \cd_{\C|\A}^\B(\calr_{\bra 0_\B} \rho_{\A\B\C} \calr_{\bra 0_\B}^\dagger)$.

Hence $Q_{\C|\B^\A \to \C|\A^\B}$ is the change of reference frame operator for states defined with respect to framed observables.

We can define the change of reference frame operator in the other direction: 
\begin{align}
    Q_{\C|\A^\B \to \C|\B^\A}(\bar \rho_{\A\C|\B} ) = \sum_i \bra{i}_\A \otimes X^i_\C  (\bar \rho_{\A\C|\B} ) X^i_\C \ket{i}_\A \otimes \ketbra{i}{i}_\B , 
\end{align}
and repeating the previous calculation gives
\begin{align}
    Q_{\C|\A^\B \to \C|\B^\A}(\bar \rho_{\A\C|\B} ) = \bar \rho_{\B\C|\A}. 
\end{align}
This explicitly proves the invertibility of the map $Q_{\C|\B^\A \to \C|\A^\B}$ with $(Q_{\C|\B^\A \to \C|\A^\B})^{-1} = Q_{\C|\A^\B \to \C|\B^\A}$.

Let us explicitly work out the states obtained for the two global states $\rho^0 = \ketbra{\psi^0}{\psi^0}$ and $\rho^1 = \ketbra{\psi^1}{\psi^1}$ used in the main text. We have:
\begin{align}
    \rho(i)_{\B\C}^{\A} = \ketbra{\psi^i}{\psi^i}_{\B\C}^\A
\end{align}
with $\ket{\psi^i}_{\B\C}^\A$ as given in~\Cref{zerowrtA,onewrtA}:
\begin{align}
    \bar \rho(0)_{\B\C}^{\A} = 
    \bar \rho(1)_{\B\C}^{\A} = \sum_k \ketbra{k}{k}_\B \otimes \bra k_\B \rho(0)_{\B\C}^{\A} \ket k_\B = \frac{1}{2} ( \ketbra{0}{0}_\B \otimes \ketbra{1}{1}_\C + \ketbra{1}{1}_\B \otimes \ketbra{1}{1}_\C), 
    \end{align}
and similarly one can calculate:
\begin{align}
    \bar \rho(0)_{\A\C}^\B =\frac 1 2 ( \ketbra{0}{0}_\A \otimes \ketbra{1}{1}_\C + \ketbra{1}{1}_\A \otimes \ketbra{1}{1}_\C ), \\
    \bar \rho(1)_{\A\C}^\B = \frac 1 2 (\ketbra{0}{0}_\A \otimes \ketbra{1}{1}_\C + \ketbra{1}{1}_\A \otimes \ketbra{1}{1}_\C ).
\end{align}

\section{Framed observables and the global invariant algebra}

The observables accessible to both $\A$ and $\B$ are given by:
\begin{align}
    (\yen_{\B\C}^\A(\cb(\ch_\B \otimes \ch_\C)) \cap \yen_{\A\C}^\B(\cb(\ch_\A \otimes \ch_\C))) \subset \cb(\ch_\A \otimes \ch_\B \otimes \ch_\C)^G,
\end{align}
where
\begin{align}
    \yen_{\B\C}^\A(\cb(\ch_\B \otimes \ch_\C)) = \{\yen_{\B\C}^\A(T_{\B\C})|T_{\B\C} \in \cb(\ch_\B \otimes \ch_\C)\}, \\
    \yen_{\A\C}^\B(\cb(\ch_\A \otimes \ch_\C)) = \{\yen_{\A\C}^\B(T_{\A\C})|T_{\A\C} \in \cb(\ch_\A \otimes \ch_\C)\} .
\end{align}

Let us define
\begin{align}
    \cf^\A_{\C|\B} = \{\yen^\A_{\B\C}(\sum_i \ketbra{i}{i}_\B \otimes C_i)|C_i \in \cb(\ch_\C)\}, 
\end{align}
which is the set of $\B$-framed observables relative to $\A$, embedded in the global invariant algebra. Similarly we define
\begin{align}
        \cf^\B_{\C|\A} = \{\yen^\B_{\A\C}(\sum_i \ketbra{i}{i}_\A \otimes C_i)|C_i \in \cb(\ch_\C)\} . 
\end{align}

Clearly  $\yen_{\B\C}^\A(\sum_i \ketbra{i}{i}_\B \otimes C_i) \in \yen_{\B\C}^{\A}(\cb(\ch_\B \otimes \ch_\C))$.  We now show that it is in $\cf_{\C|\A}^\B \subset \yen_{\C\A}^\B(\cb(\ch_\A \otimes \ch_\C))$:
\begin{align}
    \yen^\A_{\B\C} (\sum_i \ketbra{i}{i}_\B \otimes C_i) &= \sum_j \ketbra{j}{j}_\A \otimes (\sum_i \ketbra{i\oplus j}{i \oplus j}_\B \otimes X^j C_i X^j) \\
    &= \sum_{k} \ketbra{k}{k}_\B \otimes \sum_{l} X^k \ketbra{l}{l}_\A X^k \otimes X^k (X^l C_{l } X^l)X^k \\
    & = \yen^\B_{\A\C}(\sum_l \ketbra{l}{l}_\A \otimes X^l C_{l } X^l) ,
\end{align}
where $k = i \oplus j$ and $l = i$, and hence $j = l \oplus k$.

This establishes that $\cf_{\C|\B}^\A = \cf_{\C|\A}^\B$, i.e. the relativised (to Alice) framed observables of Bob are in correspondence with the relativised (to Bob) framed observables of Alice.
We note that $\yen^\A_{\B\C}(\sum_i \ketbra{i}{i}_\B \otimes C_i) = \yen^\B_{\A\C}(Q_{\C|\B^\A \to \C|\A^\B}(\sum_i \ketbra{i}{i}_\B \otimes C_i))$. We now explicitly characterise the intersection $\yen^\A_{\B\C}(\cb(\ch_\B \otimes \ch_\C)) \cap \yen^\B_{\A\C}(\cb(\ch_\A \otimes \ch_\C))$.

An element $\Omega \in \yen_{\B\C}^{\A}(\cb(\ch_\B \otimes \ch_\C))$ is of the from:
\begin{align}
    \sum_i \ketbra{i}{i}_\A \otimes (X^i_\B \otimes X^i_\C) (T_{\B\C}) (X^i_\B \otimes X^i_\C) , 
\end{align}
and similarly for an element $\Omega^{\prime} \in  \yen^\B_{\A\C}(\cb(\ch_\A \otimes \ch_\C))$.

Hence an element $\Omega \in (\yen^\A_{\B\C}(\cb(\ch_\B \otimes \ch_\C)) \cap \yen^\B_{\A\C}(\cb(\ch_\A \otimes \ch_\C)))$ must be diagonal in both $\ketbra{i}{i}_\A$ and $\ketbra{j}{j}_\B$ and therefore of the form:
\begin{align}
    \Omega = \sum_{ij} \ketbra{i}{i}_\A \otimes \ketbra{j}{j}_\B \otimes T_\C^{ij}. 
\end{align}
Moreover it is invariant ($\cg(\Omega) = \Omega$), hence
\begin{align}
    \Omega &= \sum_{ijk} \ketbra{i\oplus k}{i\oplus k}_\A \otimes \ketbra{j\oplus k}{j\oplus k}_\B \otimes X^k T_\C^{ij} X^k \\
    &= \sum_{l = i \oplus k} \ketbra{l}{l}_\A \otimes \sum_{m = i \oplus j} X^l \ketbra{m}{m}_\B X^l \otimes X^l (\sum_i X^i T_\C^{i,m \oplus i} X^i) X^l) \\
    &= \yen^\A_{\B\C}(\sum_{m} \ketbra{m}{m}_\B  \otimes T^m_\C) . 
\end{align}

Hence a generic element $ \Omega \in  \yen_{\B\C}^\A(\cb(\ch_\B \otimes \ch_\C)) \cap \yen_{\A\C}^{\B}(\cb(\ch_\A \otimes \ch_\C))$ is also in $  \cf_{\B\C}^\A = \cf_{\A\C}^\B $.

\end{document}